\documentclass[aps,prc,twocolumn,showpacs,floatfix,nofootinbib,preprintnumbers,superscriptaddress,amsmath,amssymb,longbibliography]{revtex4-1}

\usepackage{graphicx}
\usepackage{epsfig}
\usepackage{bm}
\usepackage{color}
\usepackage{float}
\usepackage{dcolumn}
\usepackage{multirow} 
\usepackage[colorlinks=true,linkcolor=blue,citecolor=blue,urlcolor=blue]{hyperref}
\usepackage{dcolumn}
\usepackage[colorinlistoftodos,prependcaption]{todonotes}

\newcommand{\beq}{\begin{equation}}
\newcommand{\eeq}{\end{equation}}
\newcommand{\beqa}{\begin{eqnarray}}
\newcommand{\eeqa}{\end{eqnarray}}

\begin{document}

\title{Coupled-cluster theory for strong entanglement in nuclei}
\thanks{This manuscript has been authored in part by UT-Battelle, LLC, under contract DE-AC05-00OR22725 with the US Department of Energy (DOE). The US government retains and the publisher, by accepting the article for publication, acknowledges that the US government retains a nonexclusive, paid-up, irrevocable, worldwide license to publish or reproduce the published form of this manuscript, or allow others to do so, for US government purposes. DOE will provide public access to these results of federally sponsored research in accordance with the DOE Public Access Plan (http://energy.gov/downloads/doe-public-access-plan).}

\author{Z.~H.~Sun} 
\affiliation{Physics Division, Oak Ridge National Laboratory, Oak Ridge, TN 37831, USA} 

\author{G.~Hagen} 
\affiliation{Physics Division, Oak Ridge National Laboratory, Oak Ridge, TN 37831, USA} 
\affiliation{Department of Physics and Astronomy, University of Tennessee, Knoxville, TN 37996, USA}

\author{T.~Papenbrock} 
\affiliation{Department of Physics and Astronomy, University of Tennessee, Knoxville, TN 37996, USA} 
\affiliation{Physics Division, Oak Ridge National Laboratory, Oak Ridge, TN 37831, USA}

\begin{abstract} 
Atomic nuclei can exhibit shape coexistence and multi-reference physics that enters in their ground states, and to accurately capture the ensuing correlations and entanglement is challenging. We address this problem by applying single-reference coupled-cluster theory based on spherical and deformed reference states and the tailored coupled-cluster method. The latter combines configuration interaction to capture static correlations with coupled-cluster theory for dynamic correlations. We compute the atomic nuclei $^{12}$C,  $^{28}$Si, and $^{56}$Ni and find that the tailored coupled-cluster method and the single-reference approach based on a deformed Hartree-Fock state yield the most accurate results.
\end{abstract}

\maketitle

\section{introduction}

In the last two decades, {\it ab initio} computations of atomic nuclei have become increasingly accurate and capable at predicting hard-to-measure properties~\cite{Navratil2004b,forssen2005,nogga2005,hagen2008,epelbaum2011,roth2012,soma2014,duguet2017,morris2018,arthuis2020,hu2021}. 
This progress is based on ideas from effective field theory~\cite{VanKolck1994,Entem2003,epelbaum2009} and the renormalization group~\cite{bogner2003,bogner2007}, and the development and application of computationally affordable methods~\cite{dickhoff2004,lee2009,tsukiyama2011,hagen2014,stroberg2016} that solve the many-body Schr\"odinger equation.  

While these advances are based on classical computing, the advent of quantum computing provides us with the opportunity to explore such computations~\cite{dumitrescu2018,stetcu2022,kiss2022}, to better understand entanglement~\cite{robin2019,beane2019,gu2023} and correlations~\cite{sun2022} in nuclei, and to think about quantum algorithms~\cite{roggero2020,choi2021,baroni2022}. 

Due to its conceptual and computational simplicity, coupled-cluster theory plays an attractive role in both classical~\cite{kuemmel1978,bishop1991,bartlett2007,hagen2014} and quantum computing~\cite{romero2017,LeeJoonho2018,stetcu2022,kiss2022}.
Approaches that combine configuration interaction and coupled-cluster theory, such as active space methods~\cite{Stolarczyk1994,Hino2006,Kinoshita2005a}, recently received attention for potential use in hybrid quantum-classical algorithms~\cite{keen2021,bauman2023}. This makes it interesting to also study them in nuclear theory. Independent of advances of quantum hardware and computing~\cite{LeeSeunghoon2022} such methods can also help us here and now in classical computing.  

Computational approaches to nuclear ground-states often start from a reference state --  a product state obtained for instance from a self-consistent mean-field calculation --  to understand the essential physics of the nucleus under consideration and to organize how to include quantum correlations for an accurate computation.
Dynamic correlations are included via particle-hole excitations of the reference state. They scale polynomially in number and their inclusion yields an extensive, i.e. proportional to mass number $A$, contribution to the binding energy~\cite{mok1996,crittenden2013,ramoscordoba2016,riveros2017}. They mix the reference state and the high-energy particle-hole excitation which reflects short-wavelength relative motion of particles~\cite{joshua2016} and short-range behavior of nuclear force.  
Single-reference methods such as  the coupled-cluster (CC) method~\cite{hagen2012d,hagen2010b,hagen2016b,morris2018} and the in-medium similarity renormalization group (IMSRG) \cite{tsukiyama2011,hergert2013b,hergert2016,stroberg2016} are designed to capture dynamic correlations with a computational effort that scales polynomially in mass number and the number of single-particle states contained in Hilbert space.  
In doubly-magic nuclei, coupled cluster with singles and doubles (CCSD) captures about $90\%$ of the correlation energy, while triples contribute about $10\%$ of the CCSD correlation energy~\cite{bartlett2007,hagen2009b}. Coupled-cluster theory with singles, doubles, and triples (CCSDT) yields about 98 to 99\% of the total correlation energy. The remaining correlation energy could be gained by including excitation beyond CCSDT; however, the 1-2\% uncertainty in the binding energy that remain at the CCSDT truncation are often smaller than the uncertainty in the employed nuclear Hamiltonians. Apart from benchmarking there is probably little need to solve approximated Hamiltonians exactly. 

Single-reference methods fail in systems that lack 
an energy gap at the Fermi surface~\cite{horoi2007,evangelista2018}. In such a case, many-body perturbation theory fails because of small energy denominators, and many-particle--many-hole excitations yield considerable contributions to the correlation energy. An alternative view is that the reference state is not the only dominant configuration and one is dealing with a multi-reference problem~\cite{lyakh2012,evangelista2018}. Then, one has to include static correlations that are due to a family of (quasi-)degenerate configurations. In nuclei, the existence of degenerate states is usually related to symmetries broken in the mean field while quasi-degenerate configurations correspond to situations of shape co-existence. Projection techniques allow one to capture the energy gain from static correlations in the case of symmetry breaking~\cite{qiu2017,qiu2018,qiu2019,frosini2022a,frosini2022b,frosini2022c,hagen2022}. In such cases, static correlations do not yield extensive contributions to the energy and usually include long wavelength physics.~\cite{peierls1957,frosini2022b,hagen2022,papenbrock2022}.

In this work, we focus on static correlations that do not result from broken symmetries. Methods such as the shell model coupled cluster~\cite{sun2018,sun2021} and valence space in-medium similarity renormalization group (VS-IM-SRG)~\cite{tsukiyama2012,stroberg2016} treat  dynamic correlations separately. Here, dynamic correlations are strongly renormalized and all correlations are capturted in a single shell of single-particle states. These methods work well in open-shell nuclei adjacent to the core with a double magic number and have been developed to calculate a large portion of the nuclear chart. However, suffering from the intruder state problem, constructing effective interaction in a valence space spanned by more than two major shells is still challenging.

Here, we follow an alternative approach and deal with the static correlation using the tailored coupled cluster (TCC) method~\cite{Stolarczyk1994,Hino2006,Kinoshita2005a,faulstich2019,vitale2020,leszczyk2022}. This approach aims to combine the simplicity of single reference coupled-cluster theory with a  multi-reference wave function. The underlying idea is as follows: In a system with two-body interactions, the correlation energy depends only on the two-particle--two-hole (2p-2h) amplitudes. While the numerical values of the amplitudes are informed by 3p-3h and higher-rank amplitudes, the latter do not enter directly the expression for the correlation energy. This makes it attractive to replace 2p-2h coupled-cluster amplitudes in the vicinity of the Fermi surface by those from methods such configuration interaction or density renormalization group and thereby include the effects of static correlations. 

The tailored coupled-cluster method partitions the single-particle basis into an active space (consisting of a smaller number of selected hole and particle states) and an external space.  The active space is selected to be large enough to reflect the static correlation of the system, in which the many-body correlations are treated exactly with full configuration interaction. The remaining correlations from the external space are captured dynamically through the standard coupled cluster method. As we will see, it is interesting to compare the results of the tailored coupled-cluster method to those from a symmetry-breaking single-reference computation.

This paper is organized as follows. In Sect.~\ref{sec:meth}, we  introduce the single-reference and tailored coupled-cluster theory. In Sect.~\ref{sec:results}, we will study the static correlations in a shell-model description of $^{56}$Ni and compare tailored and symmetry-breaking coupled-cluster methods. We also study the tailored coupled-cluster computations of $^{12}$C and $^{28}$Si using two different chiral potentials. Finally, we summarize our results in Sec.~\ref{sec:sum}.

\section{Method}
\label{sec:meth}
The intrinsic Hamiltonian consists of one-, two-, and three-body terms
\begin{equation}\label{haml}
    H=\frac{A-1}{A}
    \sum_{i=1}^A\frac{p_i^2}{2m}+\sum_{i<j}\left(v_{ij}-\frac{p_ip_j}{mA}\right)+\sum_{i<j<k}^Av_{ijk}.
\end{equation}
Here $A$ is the mass number of nucleus, $m$ is the nucleon mass, $p_i$ is the momentum of nucleon $i$, and $v_{ij}$ and $v_{ijk}$ are the two- and three-body interaction, respectively. This intrinsic Hamiltonian does not reference the center-of-mass coordinate. One normal-orders the Hamiltonian with  respect to the reference state, e.g., the Hartree Fock state 
\begin{equation}
    |\Phi_0\rangle \equiv \prod_{i=1}^A\hat{a}^\dagger_i|0\rangle \ .
\end{equation}|
The correlation energy is defined as the difference between total energy and the energy $E_0=\langle\Phi_0|H|\Phi_0\rangle$ of the reference. 
\begin{equation}
    E_{\rm{corr}}=E-E_0
\end{equation}
The normal-ordered Hamiltonian with respect to the reference reads
\begin{eqnarray}\label{no3b}
        \mathcal{H} &=& E_0+\sum_{pq}f_{pq}\left\{\hat{a}_p^\dagger \hat{a}_q \right\}+\frac14\sum_{pqrs}\Gamma_{pqrs}\left\{\hat{a}_p^\dagger \hat{a}_q^\dagger \hat{a}_s\hat{a}_r\right\}\nonumber\\
        &&+\frac1{36}\sum_{pqrstu}W_{pqrstu}\left\{\hat{a}_p^\dagger \hat{a}_q^\dagger \hat{a}_r^\dagger \hat{a}_u\hat{a}_t\hat{a}_s\right\}.
\end{eqnarray}
Here the curly brackets imply normal ordering. The operators $\hat{a}_p^\dagger$ and $\hat{a}_p$ create and annihilate a nucleon in state $|p\rangle = \hat{a}_p^\dagger|0\rangle$, respectively, and they fulfill the canonical anti-commutation relations for fermions. The matrix elements  of the one-, two-, and three-body operators are denoted as $f_{pq}$, $\Gamma_{pqrs}$, $W_{pqrstu}$, respectively. To avoid dealing with the three-body terms, one usually truncates Eq.~(\ref{no3b}) at the normal-ordered two-body level and neglects the residual three-body terms~\cite{hagen2007a,roth2012}. Thus, three-body forces enter in $E_0$, $f_{pq}$ and $\Gamma_{pqrs}$.

Coupled-cluster theory employs a similarity transformation to decouple the ground state from its particle-hole excitations 
\begin{equation}
\label{decouple}
    \langle \Phi^{a_1\cdots a_n}_{i_1\cdots i_n}|e^{-T}He^{T}|\Phi_0\rangle =0 
\end{equation}
for $n=1, \ldots A$.
Here 
\begin{equation}
 |\Phi^{a_1\cdots a_n}_{i_1\cdots i_n}\rangle \equiv   
  \hat{a}_{a_1}^\dagger \cdots \hat{a}_{a_n}^\dagger \hat{a}_{i_n}\cdots \hat{a}_{i_1} |\Phi_0\rangle
\end{equation}
is a $n$p-$n$h excitation of the reference state, and 
$T$ is the cluster operator consisting of 1p-1h, 2p-2h up to $A$p-$A$h excitations,
\begin{equation}
T=T_1+T_2+T_3+\cdots + T_A, 
\end{equation}
and 
\begin{equation}
    T_n=\frac1{(n!)^2}\sum_{\substack{a_1,a_2,\cdots,a_n\\ i_1,i_2,\cdots, i_n}}
                 t^{a_1,a_2,\cdots,a_n}_{i_1,i_2,\cdots,i_n} \hat{a}_{a_1}^\dagger \cdots \hat{a}_{a_n}^\dagger \hat{a}_{i_n}\cdots \hat{a}_{i_1} \ .
\end{equation}
In these expressions (and in what follows) we labelled hole states with indices $i,j,k,\ldots$ and particle states with $a,b,c,\ldots$. Once decoupled, the ground-state energy is calculated as 
\begin{equation}
\label{ecc}
    E_{\rm{CC}} = \langle \Phi_0|e^{-T}He^{T}|\Phi_0\rangle \ .
\end{equation}
It is important to note that -- for normal-ordered two-body Hamiltonians -- only $T_1$ and $T_2$ contribute to the energy~(\ref{ecc}), while the remaining cluster amplitudes $T_n$ with $n>2$ impact the values of $T_1$ and $T_2$ via the decoupling Eqs.~(\ref{decouple}). Thus, any inaccuracy in coupled-cluster theory could in principle be remedied by using more accurate $T_1$ and $T_2$ amplitudes. 

In the Hartree-Fock basis, the doubles yield the dominant contribution to the total binding. The estimate from many-body perturbation theory 
 \begin{equation}\label{t2est}
     T_{ij}^{ab}\approx \frac{\Gamma_{abij}}{\epsilon_a +\epsilon_b-\epsilon_i-\epsilon_j}.
 \end{equation}
indicates that the cluster amplitudes $T_{ij}^{ab}$ can be large when single-particle states that are nearly degenerate with the reference. In such situations, induced three-body terms such as 
 \begin{equation}
     H_\mathrm{3b}=\left[H,T_2\right]_{\rm{3b}}
 \end{equation}
become considerable in size. This drives higher-rank cluster amplitudes and reflects the strong entanglement in the nucleus. Single-reference coupled-cluster theory are not expected to capture such static correlations. 

\begin{figure}[!htbp]
    \includegraphics[width=0.8\linewidth]{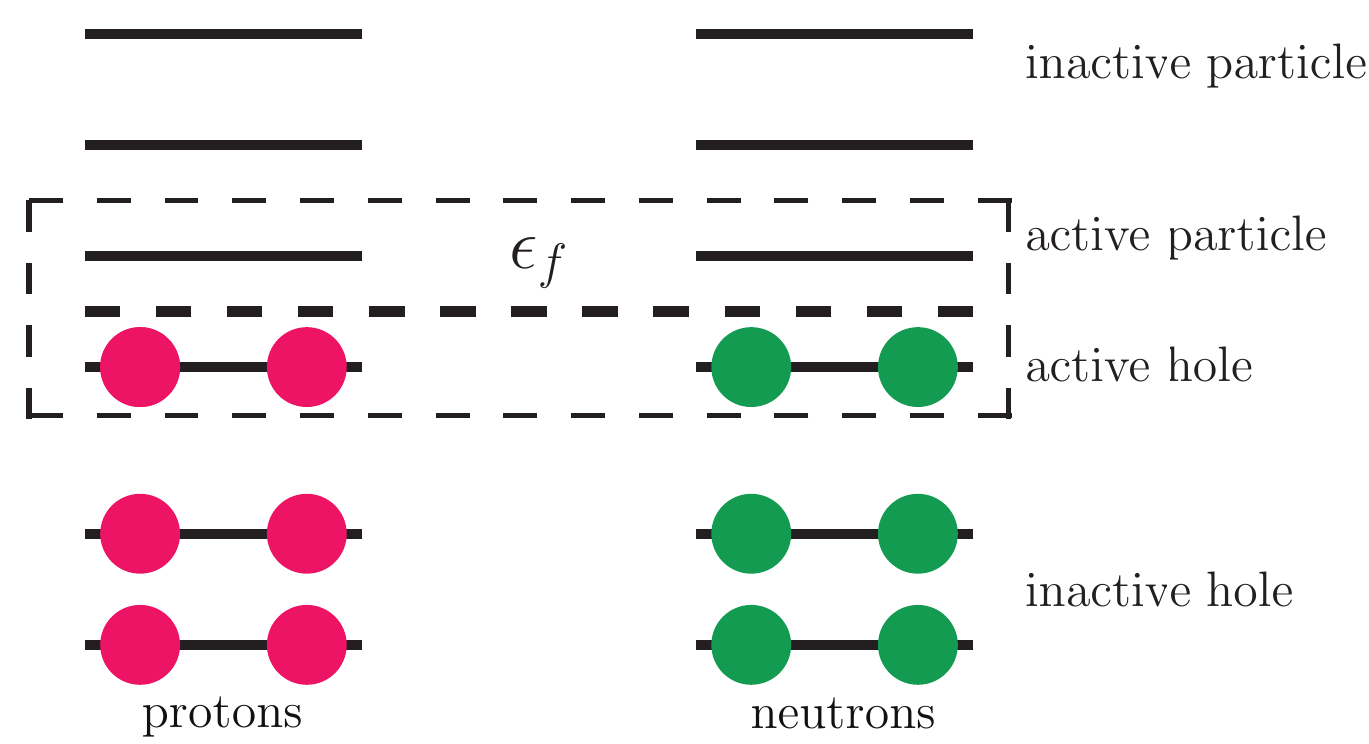}
    \caption{Partition of the single-particle orbitals into active hole, active particle, inactive hole, and inactive particle states. The orbitals in the dashed rectangle around the Fermi surface (indicated by a thick dashed line and its energy $\epsilon_f$) are part of the active space.}
    \label{fig:orbits}
\end{figure}

The tailored coupled cluster is designed to address this shortcoming. One splits the single-particle Hilbert space into an ``active'' space and ``inactive'' space. The active space is selected large enough to describe the static correlations near the Fermi surface. Figure~\ref{fig:orbits} shows the partition of the single-particle basis. The hole space is determined by a filling of shell-model orbitals, with the particle space being the complement. Near the Fermi surface a set of active particle and hole states is identified, while all other orbitals are labelled as inactive.

In tailored coupled-cluster theory, one writes the ground-state wave function as 
\begin{equation}
\label{tcc}
|\Psi_{\rm{TCC}}\rangle =e^{T_{\rm{ext}} + T_{\rm{act}}}|\Phi_0\rangle \ .
\end{equation}
Here $T_{\rm{act}}$ is the cluster operator in the active space, and all its indices refer to active particle and active hole states. $T_{\rm{ext}}$ is the cluster operator in the ``external'' space, in which at least one index belongs to an inactive particle or inactive hole state. Clearly, Eq.~(\ref{tcc}) can be rewritten as
 \begin{equation}
     |\Psi\rangle=|\Psi_{\rm{act}}\rangle\otimes |\Psi_{\rm{ext}}\rangle \ .
 \end{equation}

The amplitudes $T_{\rm{act}}$ are obtained from the full configuration interaction wave function in the active space, which we expand as
\begin{equation}
\label{FCIact}
    |\Psi_{\rm{act}} \rangle = C_0|\Phi_0\rangle +C^a_i|\Psi_i^a\rangle + C^{ab}_{ij}|\Psi_{ij}^{ab}\rangle+\cdots \ .
\end{equation}
Here $C_0$, $C^a_i$, etc. are the coefficients of the reference state, its 1p-1h, 2p-2h, and up to $A$p-$A$h excitations, respectively, and summation over repeated indices is implied. The indices $a,b,i,j$ denote active particle and active hole states.
We can also expand
\begin{eqnarray}
\label{CCact}
      |\Psi_{\rm{act}} \rangle &=& e^{T_{\rm{act}}}|\Phi_0\rangle\nonumber\\
      &=& \left[1+T_1+\left(T_2 + {1\over 2} T_1^2\right)+\cdots\right]|\Phi_0\rangle \ .  
\end{eqnarray}
Here, we have dropped the subscript ``act'' from the cluster amplitudes on the right-hand-side of Eq.~(\ref{CCact}), and it is implied that all indices refer to active single-particle states.  Working in the CCSD approximation we get the active coupled-cluster amplitudes $T_{\rm{act}}$ by comparing Eqs.~(\ref{FCIact})  and (\ref{CCact}) giving, 
\begin{equation}
    T^a_i=\frac{C^a_i}{C_0} \ ,
\end{equation}
and 
\begin{equation}
    T^{ab}_{ij}=\frac{C^{ab}_{ij}}{C_0} - \frac{C^a_iC^b_j-C^a_jC^b_i}{C_0^2} \ .
\end{equation}
This can easily be extended to also include triples in the active space in CCSDT. In these equations it is again implied that all indices of the cluster amplitudes refer to active-space orbitals. We note that the CCSD energy $ \langle \Phi_0| \left(He^{T_{\rm{act}} }\right)_c|\Phi_0 \rangle $ equals the full configuration interaction energy in the active space.  

The external cluster operator $T_\mathrm{ext}$ is calculated by solving the usual coupled-cluster equations while keep $T_{\rm{act}}$ fixed. On the CCSD level the decoupling equation for the external space reads
\begin{equation}
\label{tcca}
    \langle \Phi^a_i| \left(He^{T_{\rm{act}} + T_{\rm{ext}}}\right)_c|\Phi_0 \rangle = 0 \ , 
\end{equation}
and 
\begin{equation}
\label{tccb}
    \langle \Phi^{ab}_{ij}| \left(He^{T_{\rm{act}} + T_{\rm{ext}}}\right)_c|\Phi_0 \rangle = 0  \ .
\end{equation}
Here, it is understood that at least one of the displayed indices refers to an inactive orbital.  
Once the $T_{\rm{ext}}$ is determined, the energy is calculated as in eq.~(\ref{ecc}) but with $T=T_\mathrm{act}+ T_\mathrm{ext}$, i.e.
\begin{equation}
E_{\rm{TCCSD}}=
         \langle \Phi_0| \left(He^{T_{\rm{act}} + T_{\rm{ext}}}\right)_c|\Phi_0 \rangle \ .
\end{equation}
Here, the subscript $c$ implies that  only connected diagrams are included on the right-hand side. 
Equations~(\ref{tcca}) and (\ref{tccb}) show that the active space feeds back into the amplitudes of the external space, but that the latter does not impact the former. This is a shortcoming of tailored coupled-cluster method, and to remedy this would require a large active space such that both static and dynamic correlations of the exact wave function is accurately described~\cite{anna2012}.

In single-reference coupled-cluster theory the triples  (3p-3h) amplitudes yield about 10\% of the CCSD energy~\cite{bartlett2007,hagen2009b}. 
Triples corrections to the tailored CCSD approximation are only considered in the external space~\cite{Lyakh2011}. As a result, the triples correction is smaller than in the standard single-reference approach. The triples amplitudes solve
\begin{equation}
\label{tccb1}
    \langle \Phi^{abc}_{ijk}| \left(He^{T_{\rm{act}} + T_{\rm{ext}}+T_3}\right)_c|\Phi_0 \rangle = 0 \ .
\end{equation}
Here, $T_{\rm{ext}}$ contains up to 2p-2h operators.

The tailored coupled-cluster method takes a cluster operator that describes the static correlation externally from the full configuration interaction method. In treating the dynamic correlations in the external space only, there is no feedback to the active space. To get an accurate result, one needs to select the active space large enough not only to guarantee the static correlations are captured, but the wave function of the active space should also reflect the "true" wave function. The shortcoming of neglecting the coupling between the active and external space has been studied by the extended active space coupled cluster method~\cite{anna2012}. Here one first calculates the full configuration interaction wave function in a large active space, and subsequently a smaller set of active orbitals is used in the decoupling of external space in the tailored coupled-cluster method. In this way there is indirect feedback from the external to the active space. 

\section{Results}
\label{sec:results}

\subsection{Shell-model calculations}
The shell-model provides us with a framework to investigate static correlations and to what extent the tailored coupled-cluster method is able to capture them. We consider the $0f1p$ shell and the  GXPF1A effective Hamiltonian~\cite{honma2005}. The model space is spanned by the $f_{7/2}$, $p_{3/2}$,$f_{5/2}$, and $p_{1/2}$ orbitals. For $^{56}$Ni, the valence protons and neutrons fill the $f_{7/2}$ shells. To control the amount of static correlations, we follow~\textcite{horoi2007}, and shift the single-particle energies of the  $p_{3/2}$, $f_{5/2}$, and $p_{1/2}$ orbitals relative to the $f_{7/2}$ shell by a constant amount $\Delta G$ from the GXPF1A parameters. (The original GXPF1A interaction has $\Delta G=0$.) A positive value of $\Delta G$ increases the shell gap and reduces static correlations, while a negative $\Delta G$ reduces the shell gap and leads to a system with stronger static correlations. \textcite{horoi2007} showed that single-reference coupled-cluster theory fails to reproduce the ground state energy for sufficiently large negative values of $\Delta G$. This is expected because a decrease of the shell gap introduces multi-reference physics and destabilizes the Fermi surface.

\begin{figure}[!htp]
    \includegraphics[width=0.99\linewidth]{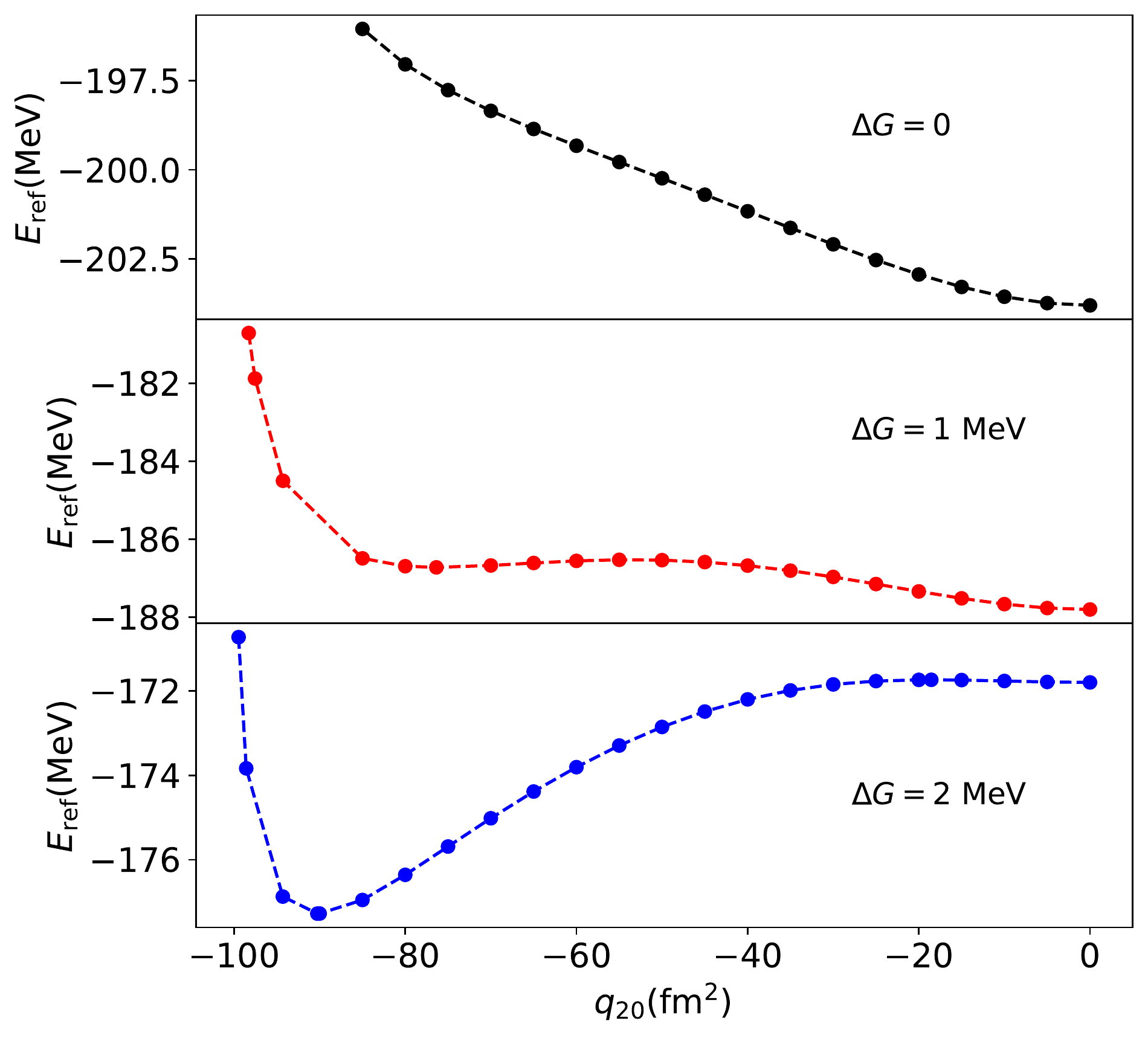}
    \caption{Energy of axially symmetric Hartree-Fock reference states as a function of the quadrupole moment $q_{20}$ for $^{56}$Ni computed with the GXPF1A interaction (bottom) and for a modified interaction where the single-particle orbitals above the $f_{7/2}$ shell are shifted by $\Delta G$.}
    \label{fig:ni56hf}
\end{figure}

We performed Hartree-Fock calculations for the Hamiltionian
\begin{equation}
\label{Hprime}
 H'(\lambda) = H+\lambda Q_{20}   
\end{equation}
where $Q_{20}$ is the quadrupole operator. 
This yields axially symmetric reference states. Once the reference was determined, we computed the energy $E_\mathrm{ref}\equiv\langle\Phi_0(\lambda)|H|\Phi_0(\lambda)\rangle$ and also the
quadrupole moment $q_{20}\equiv\langle\Phi_0(\lambda)|Q_{20}|\Phi_0(\lambda)\rangle$, and plotted $E_\mathrm{ref}$ as a function of $q_{20}$. 
The resulting energy surfaces are shown in Fig.~\ref{fig:ni56hf} for $\Delta G=0$, $-1$ and $-2$~MeV (from bottom to top). We see that there is a quantum phase transition in $^{56}$Ni from a spherical to a deformed ground state as $\Delta G$ is decreased from 0 to $-2$~MeV.

\begin{figure}[!htbp]
    \includegraphics[width=0.99\linewidth]{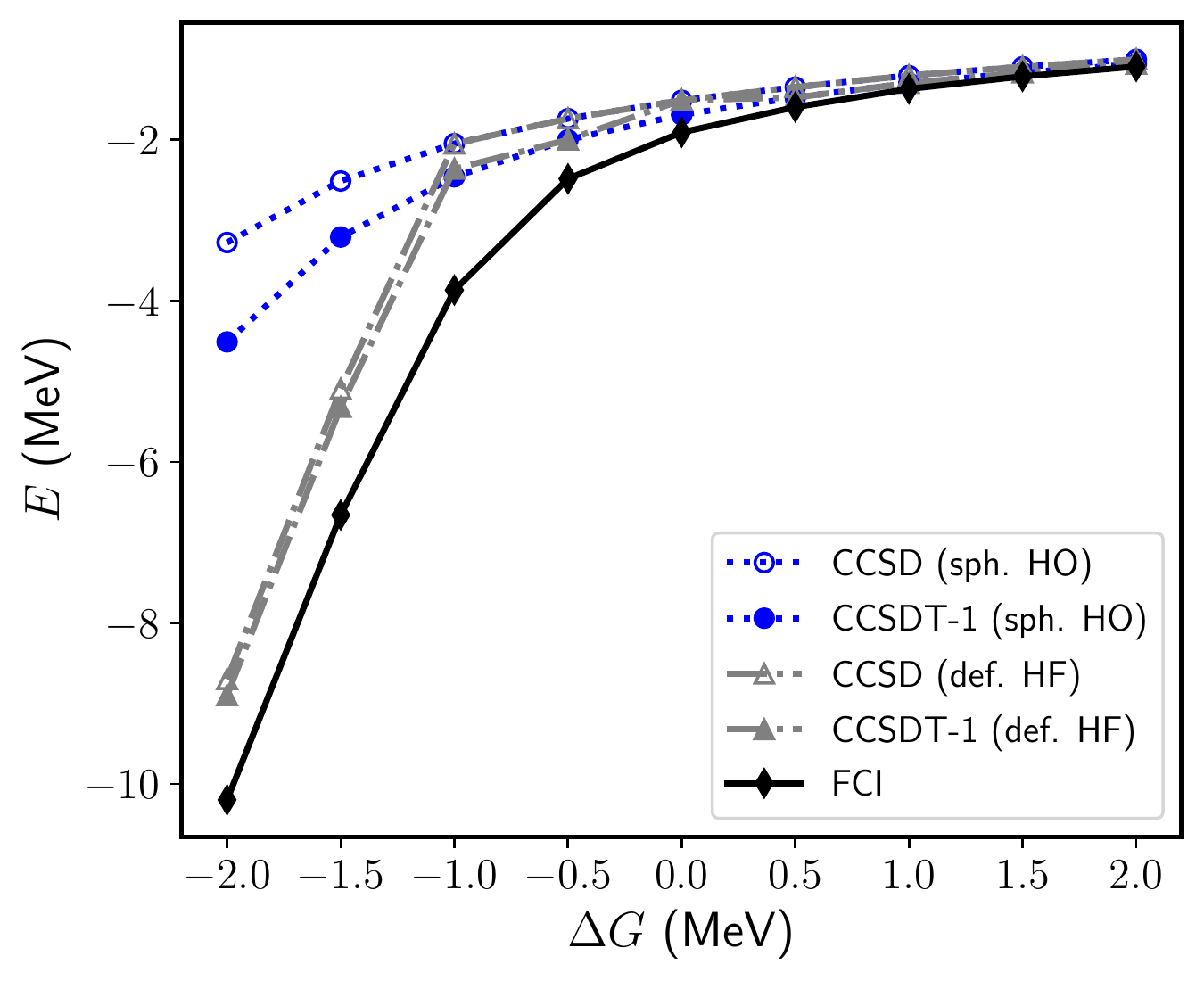}
    \caption{Ground-state energy of $^{56}$Ni, based on the GXPF1A interaction, as a function of the shift $\Delta G$ between the $f_{7/2}$ orbital and the unoccupied orbitals in this nucleus. The original interaction has $\Delta G=0$. Results from a spherical (sph.) and deformed (def.) reference state based on the CCSD and CCSDT-1 approximations are compared to full configuration interaction (labelled FCI).}
    \label{fig:ni56cc}
\end{figure}

We performed three types of  coupled-cluster computations, using  (i) a spherical, (ii) an axially symmetric reference, and (iii) tailored active spaces.  First, we take the filled $f_{7/2}$ shell as the spherical reference state and perform CCSD and coupled-cluster singles-and-doubles and leading-order iterative triples  (CCSDT-1)~\cite{lee1984} computations. The resulting ground-state  energies are shown as squares and circles, respectively in Fig.~\ref{fig:ni56cc} and can be compared to the exact result from full configuration interaction. As already documented by  \textcite{horoi2007}, spherical single-reference coupled-cluster theory fails for sufficiently large negative values of $\Delta G$. Second, we perform CCSD and CCSDT-1 computations using the axially symmetric reference state of the Hamiltonian~(\ref{Hprime}) using $\lambda = 0$. For $\Delta G \gtrsim -1$~MeV, the reference is spherical, while smaller values yield an axially symmetric (and deformed) reference. The results are shown as triangles in Fig.~\ref{fig:ni56cc}. We see that the transition region $\Delta G\approx -1$~MeV is not well described by single-reference methods, while coupled-cluster theory (based on an axially-symmetric reference state) becomes more accurate in the well deformed regime. For $\Delta G=-2$ a clear minimum develops on the energy surface (see Fig.~\ref{fig:ni56hf}) and the deformed Hartree-Fock state captures 5~MeV more in binding than the spherical one; the deformed CCSD captures about another 3~MeV. The total binding from the deformed CCSDT-1 is about 1~MeV away from the full configuration interaction. We estimate that the discrepancy to full configuration interaction is mainly due to the broken rotational symmetry. The correlation energy from deformed and spherical coupled cluster is 3.4 and 4.5~MeV, respectively. Thus, the symmetry breaking mean-field captures significantly more dynamic correlations.

\begin{figure}[!htbp]
    \includegraphics[width=0.99\linewidth]{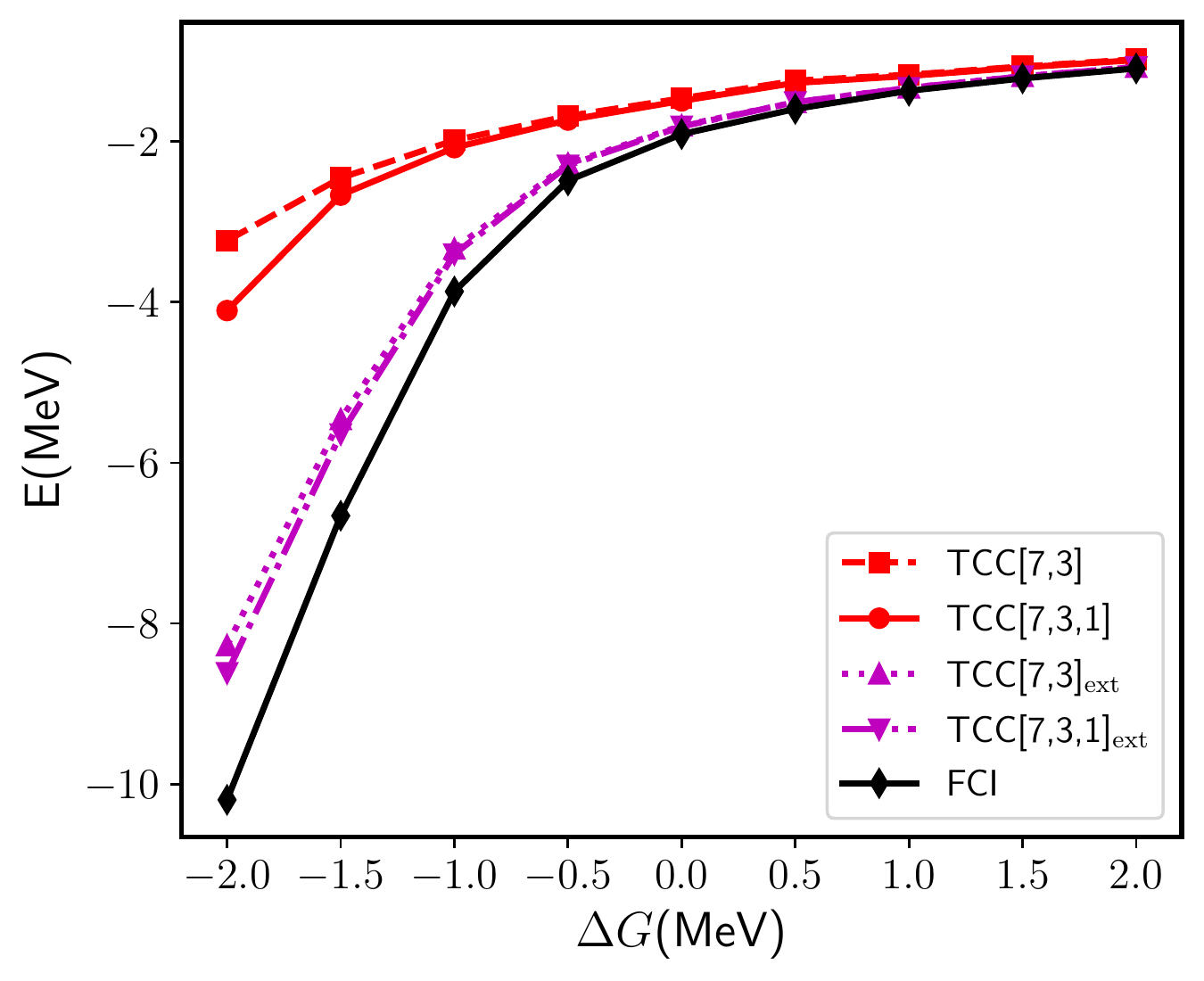}
    \caption{Tailored coupled cluster (labelled TCC) computations (identified by the set of quantum numbers $2j$, in square brackets, contained in the active space) of the ground-state energy of $^{56}$Ni, calculated with GXPF1A interaction, as a function of the shift $\Delta G$ between the $f_{7/2}$ orbital and the unoccupied orbitals in this nucleus. The original interaction has $\Delta G=0$. Also shown are computations using the extended (ext) tailored coupled-cluster method. Results are compared to full configuration interaction (labelled FCI).}
    \label{fig:ni56tailored}
\end{figure}

Third and finally, we perform tailored coupled-cluster computations. We select the $f_{7/2}$ and $p_{3/2}$ as the active space for both protons and neutrons and find that the results are very close to the CCSD computations based on the spherical reference, see Fig.~\ref{fig:ni56tailored}. Next, we extend the active space to also include the $p_{1/2}$ orbital. As shown in Fig.~\ref{fig:ni56tailored}, this leads only to a modest improvement. 
Clearly, the shell-model wave functions are highly correlated, and full configuration interaction calculations based on the  $f_{7/2}$ and $p_{3/2}$ (or the $f_{7/2}$, $p_{1/2}$, and $p_{3/2}$) orbitals do not yield accurate wave-function amplitudes. As a final check, we employ the extended tailored coupled-cluster method~\cite{anna2012}. Here, we use the exact amplitudes from a full configuration interaction calculation and then choose active spaces consisting of the $f_{7/2}$ and $p_{3/2}$ (and the $f_{7/2}$, $p_{1/2}$, and $p_{3/2}$) orbitals. In these cases tailored coupled-cluster yields significantly more accurate results, see Fig.~\ref{fig:ni56tailored}.

Overall we see that the single-reference coupled-cluster computations that are based on a deformed (axially-symmetric) Hartree-Fock state are quite accurate when compared to those from tailored coupled-cluster theory when the shape of the nucleus is spherical or deformed. Of course, the former lack rotational invariance for small values of $\Delta G$, and this might be the price to pay for computational and conceptual simplicity.

\subsection{Calculations of $^{12}$C from chiral interactions}

Single-reference coupled-cluster theory is efficient for nuclei with a significant shell gap such as closed-shell nuclei or well-deformed nuclei (which exhibit large shell gaps in a Nilsson diagram). The most challenging nuclei are those with filled subshells that lack a clearly developed shell gap. Examples are $^{12}$C and $^{28}$Si. These spherical nuclei are somewhat soft to deformation and therefore difficult to compute accurately with single-reference methods. This makes it interesting to use tailored coupled-cluster theory.

We start with $^{12}$C, where the last four protons and four neutrons occupy the $p_{3/2}$ orbital. However, the shell gap between $p_{3/2}$ and $p_{1/2}$ is small, and $^{12}$C is not a closed-shell nuclei. Its structure is interesting and complicated~\cite{navratil2000,epelbaum2012,lovato2013}. 

\begin{figure}[!htbp]
    \includegraphics[width=0.99\linewidth]{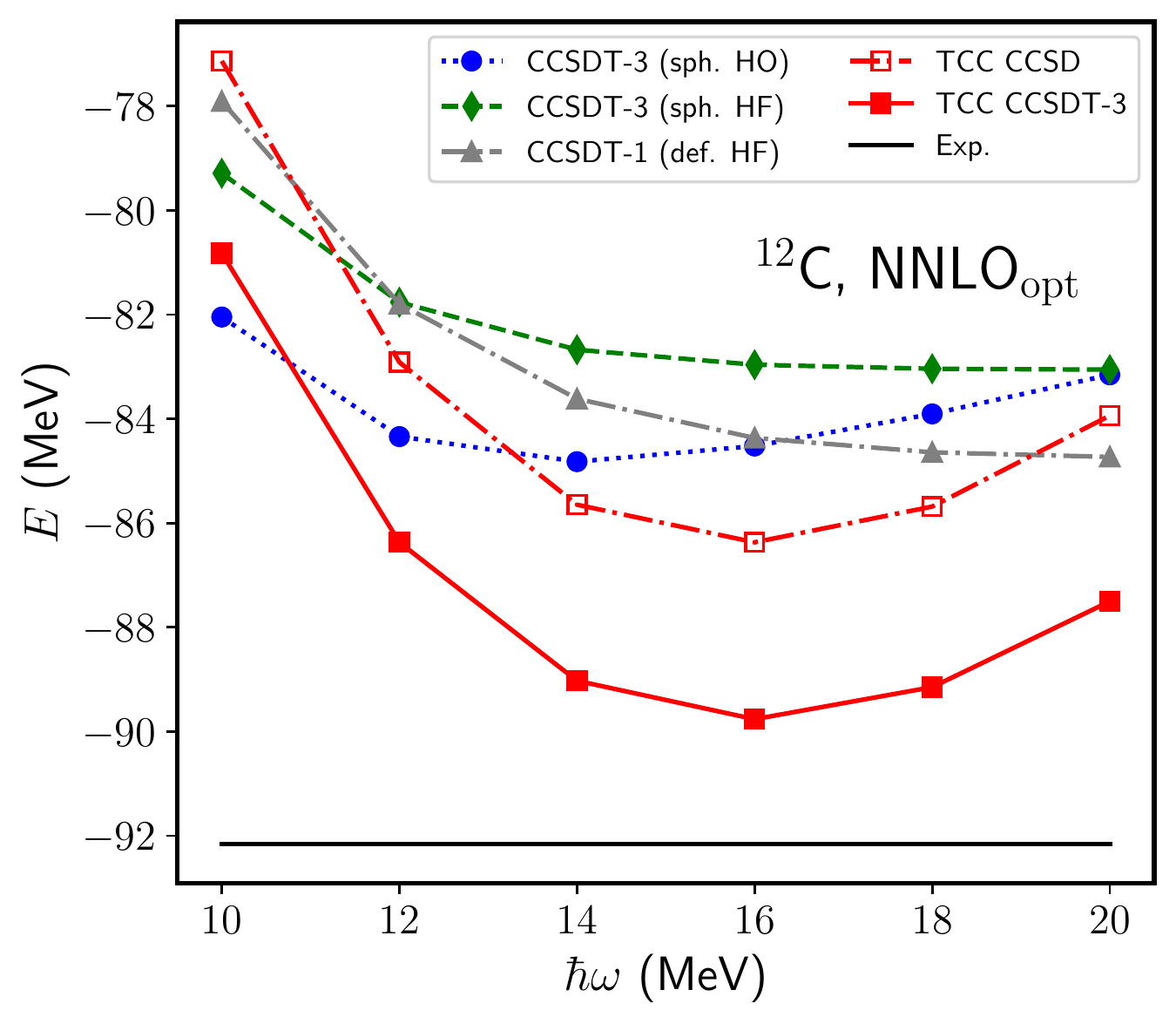}
    \caption{Ground-state energy of $^{12}$C calculated with single-reference coupled cluster methods based on reference states in the spherical harmonic oscillator harmonic-oscillator (sph. HO) basis, the spherical Hartree-Fock (sph. HF) basis, and the axially symmetric, deformed, Hartree-Fock basis (def. HF), using CCSDT-3 and CCSDT-1 approximations. Also shown are the tailored coupled-cluster calculations (labelled TCC) and the experimental data. The two-body interaction is NNLO$_{\rm{opt}}$.}
    \label{fig:c12_nnloopt}
\end{figure}

We compute $^{12}$C using the NNLO$_{\rm{opt}}$ nucleon-nucleon interaction~\cite{ekstrom2013} and an oscillator basis with frequency $\hbar\omega$ and highest single-particle energy $N_{\rm max}\hbar\omega$ (omitting the zero-point energy). We use $N_{\rm{max}}=12$ unless stated otherwise. In a first approach, we employ a reference state that fills the harmonic oscillator orbitals up to and including the $p_{3/2}$ orbitals and perform coupled-cluster computations using the CCSDT-3 approximation~\cite{urban1985}. The results are shown (as a function of the oscillator spacing $\hbar\omega$) in Fig.~\ref{fig:c12_nnloopt} and compared to the experiment. The interaction NNLO$_{\rm{opt}}$ is presumably accurate for light nuclei and this suggests that about 7~MeV of binding energy are still lacking at the triples result. We also performed calculations where the spherical Hartree-Fock state serves as the reference. As can be seen in Fig.~\ref{fig:c12_nnloopt} the results are close to those of the harmonic oscillator basis. Finally, we performed CCSDT-1 computations based on an axially symmetric (and deformed) Hartree-Fock reference. These results yield a ground state that is close to the one in the spherical harmonic-oscillator basis and about 2~MeV lower than the CCSDT-3 results using the spherical Hartree-Fock basis. 

For the tailored coupled-cluster calculations the active space is spanned by the $p_{3/2}$, $sd$, and $pf$ shell orbitals in the harmonic oscillator basis. The large active space is motivated by the need to potentially include some effects of deformation. Thus, we need to include the $pf$ shell as it has the same parity as the occupied $p_{3/2}$.  
The dimension of Hamiltonian matrix in the active space is about $2\time 10^8$, which is almost the limit of our current implement of the code. 
The results at the singles and doubles, and at the triples approximation -- also shown in Fig.~\ref{fig:c12_nnloopt} -- exhibit a significant gain in 
the ground-state energy. This is re-assuring and points to the value of the tailored coupled-cluster method for this nucleus. 
We also performed tailored coupled-cluster calculations in the Hartree-Fock basis. This, however, was not as successful as for the harmonic oscillator basis, because the tailored coupled-cluster results did not improve over the CCSDT-3  results. We do not understand why this is so.

\begin{figure}[!htbp]
    \includegraphics[width=0.99\linewidth]{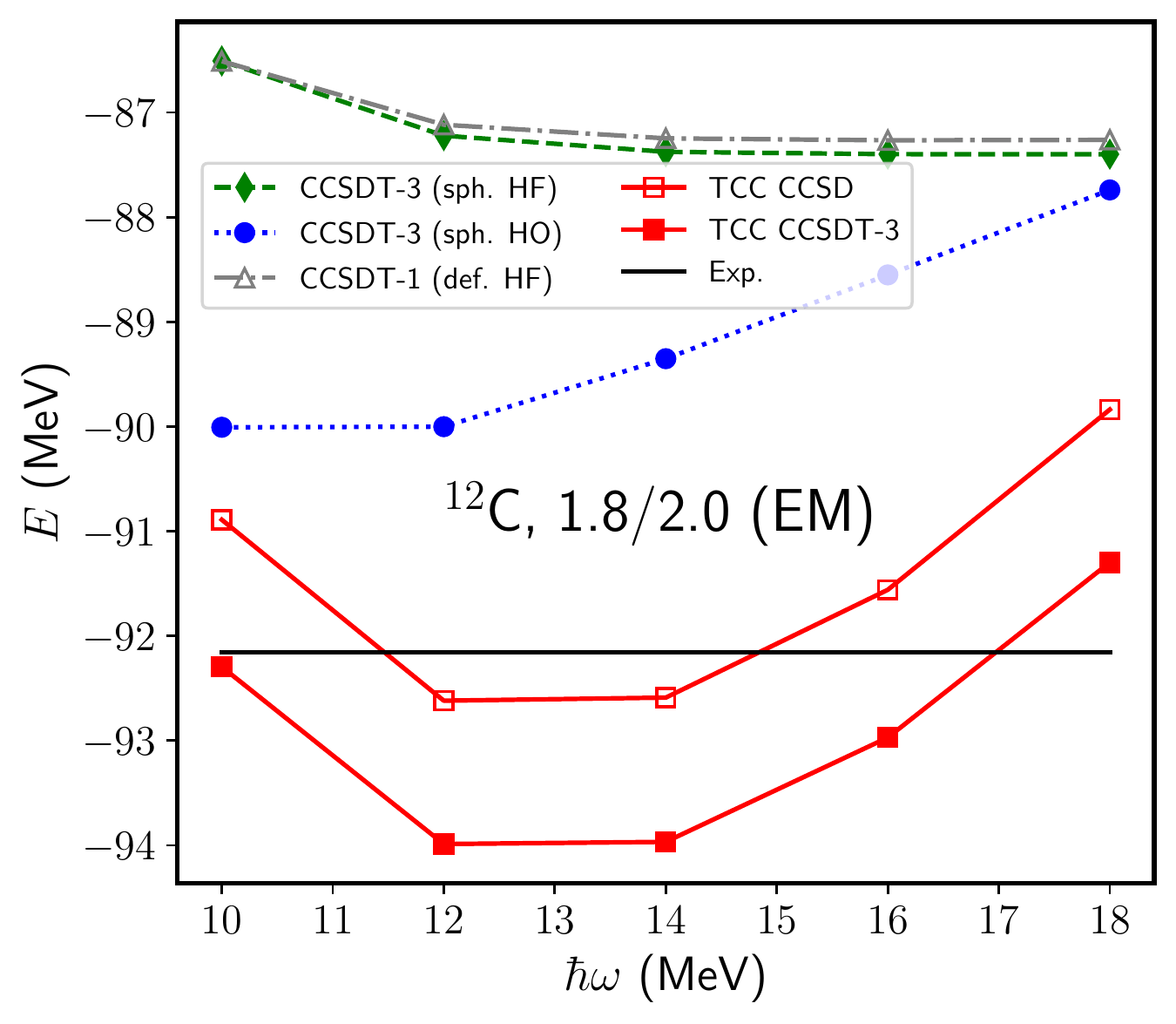}
    \caption{Ground-state energy of $^{12}$C calculated with single-reference coupled cluster methods based on reference states in the spherical harmonic-oscillator (sph. HO) basis, the spherical Hartree-Fock (sph. HF) basis, and the axially symmetric, deformed, Hartree-Fock basis (def. HF), using CCSDT-3 and CCSDT-1 approximations. Also shown are the tailored coupled-cluster calculations (labelled TCC) and the experimental data. The employed chiral interaction is 1.8/2.0~(EM) from Ref.~\cite{hebeler2011}.}
    \label{fig:c12_heb}
\end{figure}

We turn to calculations with three-nucleon forces and employed the chiral interaction  1.8/2.0~(EM) from Ref.~\cite{hebeler2011}. For the coupled-cluster calculations we want to employ the normal-ordered 
two-body approximation~\cite{hagen2008,roth2012}. While this is straight forward in the Hartree-Fock basis, the calculations in the harmonic-oscillator basis require more thought and care; a naive application of the  normal-ordered two-body approximation would introduce a strong $\hbar\omega$-dependence of the results. To avoid this we perform a spherical Hartree-Fock calculation, and normal-order the Hamiltonian with respect to the Hartree-Fock reference. We employ the normal-ordered two-body approximation, discard the normal-ordered residual three-nucleon terms, and store the normal-ordered one- and two-body matrix elements of the Hamiltonian.   Coupled-cluster calculations in the Hartree-Fock basis then use these matrix elements. For the coupled-cluster calculations in the harmonic oscillator basis
we re-normal-order the stored matrix elements of the one- and two-body Hamiltonian with respect to the true vacuum and transform to the harmonic oscillator basis~\cite{djarv2021,frosini2021}. This then yields one- and two-body matrix elements that contain the relevant contributions of the three-nucleon force.
Applications of this approach in the no-core shell-model verified that the impact on the center-of-mass separation is controlled~\cite{djarv2021}.
Figure~\ref{fig:c12_heb} shows the ground state energy of $^{12}$C calculated with spherical coupled-cluster theory in the Hartree-Fock and the harmonic oscillator basis, using the CCSDT-3 approximation. 

We also performed similar calculations in a deformed (yet axially symmetric) Hartree-Fock basis, using a quadrupole constraint to map out the energy surface. We found an oblate deformed minimum which we used as the Hartree-Fock reference. This agrees with the dominant ``triangle'' configuration of $\alpha$ clusters found in Ref.~\cite{epelbaum2012}. In these calculations we performed the normal-ordered two-body approximation in the deformed reference state. The CCSDT-1 results in this basis are very close to the CCSDT-3 results in the spherical harmonic-oscillator basis.

The tailored coupled-cluster calculations are performed in the harmonic-oscillator basis.  The active space is spanned by $p_{3/2}$ shell and the $sd$ and $pf$ shells. We verified that our calculations are converged  at $N_{\rm{max}}=12$. The tailored coupled-cluster results depend on the harmonic-oscillator spacing and they overbind the nucleus somewhat. This latter observation is consistent with similar results from quantum chemistry, where small active spaces in multi-reference methods tend to overestimate the role of static correlations~\cite{evangelista2018}. We note that the tailored coupled-cluster results deviate about as much from experiment as the coupled-cluster results obtained in the spherical harmonic-oscillator basis.

\subsection{Calculations of $^{28}$Si from chiral interactions}

\begin{figure}[!htbp]
    \includegraphics[width=0.99\linewidth]{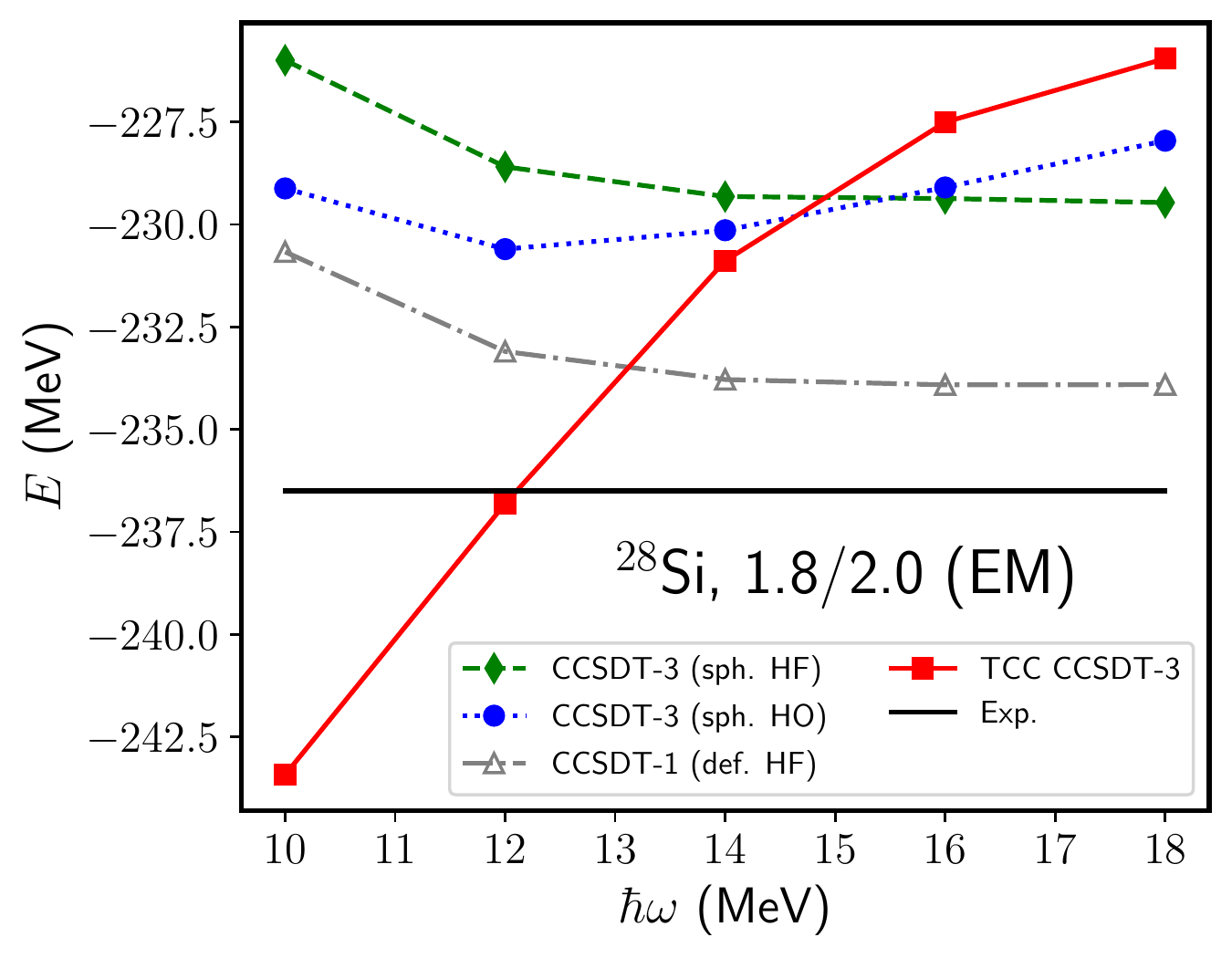}
    \caption{Ground-state energy of $^{28}$Si calculated with single-reference coupled cluster methods based on reference states in the spherical harmonic-oscillator (sph. HO) basis, the spherical Hartree-Fock (sph. HF) basis, and the axially symmetric, deformed, Hartree-Fock basis (def. HF), using CCSDT-3 and CCSDT-1 approximations. Also shown are the tailored coupled-cluster calculations (labelled TCC) and the experimental data. The employed chiral interaction is 1.8/2.0~(EM) from Ref.~\cite{hebeler2011}.}
    \label{fig:si28_heb}
\end{figure}

The nucleus $^{28}$Si, with filled $d_{5/2}$ shells in the shell model, is another challenging nucleus. We employ the 1.8/2.0(EM) chiral potential in a model space of $N_{\rm{max}}=12$, and show our results in Fig.~\ref{fig:si28_heb}. The single-reference coupled-cluster calculations are again based on a spherical harmonic-oscillator state, and on spherical and deformed Hartree-Fock states; from these, the deformed reference yields most accurate results (and a gain of about 3~MeV in the binding energy). We see that $^{28}$Si differs in this regard from $^{12}$C. It is somewhat more deformed in its ground state. 

In our tailored coupled-cluster calculation, the active space is spanned by the $sd$ and $f_{7/2}$ shells, filled with six protons and six neutrons. This requires us to perform large-scale shell-model calculations. The results are clearly not converged and strongly depend on the oscillator frequency. We understand the lack of convergence as follows: To capture quadrupole deformation from the $d_{5/2}$ shell, we would need to also include the $0g1d2s$ shell in the active space (because the quadrupole operator couples the $0d1s$ and $0g1d2s$ shells), but this is not possible with current computing resources (the shell-model dimension is beyond $10^9$). We also performed similar calculation by using an active space spanned by the $p$ and $sd$ shells,  with 12 protons ans 12 neutrons in the active space. This did not significantly change the results. The active space we employ -- though large from its computational demands -- is too small  to capture the deformation.

\begin{figure}[!htbp]
    \includegraphics[width=0.99\linewidth]{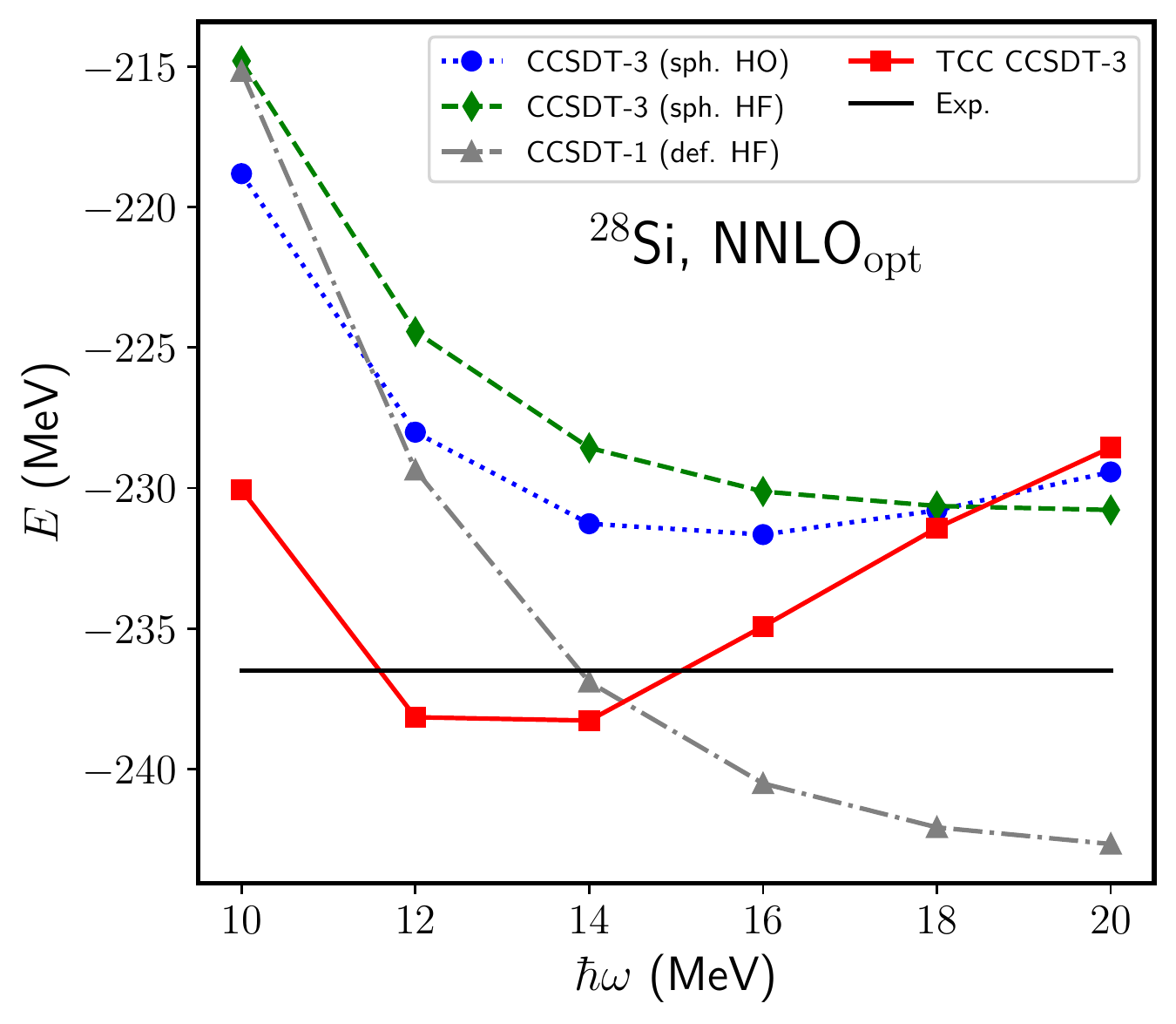}
    \caption{Ground-state energy of $^{28}$Si calculated with single-reference coupled cluster methods based on reference states in the spherical  harmonic-oscillator (sph. HO) basis, the spherical Hartree-Fock (sph. HF) basis, and the axially symmetric, deformed, Hartree-Fock basis (def. HF), using CCSDT-3 and CCSDT-1 approximations. Also shown are the tailored coupled-cluster calculations (labelled TCC) and the experimental data. The two-body interaction is NNLO$_{\rm{opt}}$.}
    \label{fig:si28_nnloopt}
\end{figure}

For a simpler calculation we studied $^{28}$Si with the NNLO$_{\rm{opt}}$ interaction, using the active space consisting of the $sd$ and $f_{7/2}$ shells. Results are shown in Fig.~\ref{fig:si28_nnloopt}.
The single-reference coupled-cluster calculations indicate that the nucleus exhibits stronger deformation than for the  1.8/2.0(EM) interaction, because calculations based on a deformed reference gain about 10~MeV in binding. The NNLO$_{\rm{opt}}$ potential is probably not as accurate in this region of the nuclear chart as the  1.8/2.0~(EM) interaction, and -- based on our experience with single-reference coupled-cluster calculations -- it overbinds the $^{28}$Si nucleus. The tailored coupled-cluster results are better converged than for the 1.8/2.0~(EM) interaction, although the $\hbar\omega$-dependence is still strong. We attribute the difference between the tailored coupled-cluster results and the calculations with the deformed Hartree-Fock reference on the too-small active space that does not capture deformation. 

\section{Summary}
\label{sec:sum}
Single-reference coupled-cluster calculations become challenging in nuclei that are neither strongly deformed (i.e. exhibiting rigid-rotor-like ground-state rotational bands) nor spherical, e.g. for $^{12}$C, $^{28}$Si, and $^{56}$Ni. In such situations a combination of full configuration interaction and coupled-cluster methods seems promising, and the prospect of hybrid quantum-classical computing makes this interesting to study. 

We find that tailored coupled cluster methods can be alternatives to single-reference methods in certain nuclei. They typically gain energy compared to spherical single-reference methods and keep the symmetry of the wave function intact. For computations of ground-state energies, symmetry-broken single-reference approaches are quite competitive. While they require symmetry projection (in contrast to tailored coupled-cluster), they exhibit better convergence with respect to  model-space parameters than tailored coupled cluster methods.
   
\begin{acknowledgments}
This work is supported by the Quantum Science Center, a National Quantum Information Science Research Center of the U.S.~Department of Energy. It is also supported by the U.S. Department of Energy, Office of Science, Office of Nuclear Physics under Award Numbers DEFG02-96ER40963, DE-SC0018223 (SciDAC-4 NUCLEI), and the SciDAC-5 NUCLEI collaboration.
Computer time was provided by the Innovative and Novel Computational Impact on Theory and Experiment (INCITE) program. This research used resources of the Oak Ridge Leadership Computing Facility located at Oak Ridge National Laboratory, which is supported by the Office of Science of the Department of Energy under Contract No. DE-AC05-00OR22725.
\end{acknowledgments}

\bibliography{references}
\end{document}